\documentclass[preprint,prd,aps,nofootinbib]{revtex4}
\usepackage{amsmath}
\usepackage{psfig}
\newcommand{\be}{\begin{equation}}
\newcommand{\ee}{\end{equation}}
\newcommand{\ba}{\begin{eqnarray}}
\newcommand{\ea}{\end{eqnarray}}
\newcommand{\bc}{\begin{center}}
\newcommand{\ec}{\end{center}}

\begin{document}
\begin{center}
\bibliographystyle{article}

{\Large \textsc{Gravitational amplitudes in black-hole evaporation:
the effect of non-commutative geometry}}

\end{center}
\vspace{0.4cm}


\date{\today}

\author{Elisabetta Di Grezia,$^{1}$
\thanks{Electronic address: digrezia@na.infn.it}
Giampiero Esposito,$^{1,2}$ \thanks{
Electronic address: giampiero.esposito@na.infn.it}
Gennaro Miele,$^{2,1}$
\thanks{Electronic address: gennaro.miele@na.infn.it}}

\affiliation{${\ }^{1}$Istituto Nazionale di Fisica Nucleare,
Sezione di Napoli,\\
Complesso Universitario di Monte S. Angelo, Via Cintia, Edificio N', 80126
Napoli, Italy\\
${\ }^{2}$Dipartimento di Scienze Fisiche, Complesso Universitario di Monte
S. Angelo,\\
Via Cintia, Edificio N', 80126 Napoli, Italy}

\begin{abstract}
Recent work in the literature has studied the quantum-mechanical decay of
a Schwarzschild-like black hole, formed by gravitational collapse, into
almost-flat space-time and weak radiation at a very late time. The relevant
quantum amplitudes have been evaluated for bosonic and fermionic fields,
showing that no information is lost in collapse to a black hole. On the
other hand, recent developments in noncommutative geometry have shown
that, in general relativity, the effects of noncommutativity can be taken
into account by keeping the standard form of the Einstein tensor on the
left-hand side of the field equations and introducing a modified
energy-momentum tensor as a source on the right-hand side. The present
paper, relying on the recently obtained noncommutativity effect on a
static, spherically symmetric metric, considers from a new perspective the
quantum amplitudes in black hole evaporation. The general relativity
analysis of spin-2 amplitudes is shown to be modified by a multiplicative
factor $F$ depending on a constant non-commutativity parameter and on the
upper limit $R$ of the radial coordinate. Limiting forms of $F$ are
derived which are compatible with the adiabatic approximation here
exploited. Approximate formulae for the particle emission rate are also
obtained within this framework.
\end{abstract}

\maketitle
\bigskip
\vspace{2cm}

\section{Introduction}

Theoretical research in black hole physics has witnessed,
over the last few years, an impressive
amount of new ideas and results on at least four main areas:
\vskip 0.3cm
\noindent
(i) The problem of information loss in black holes, after the suggestion
in Ref. \cite{Hawk05} that quantum gravity is unitary and information is
preserved in black hole formation and evaporation.
\vskip 0.3cm
\noindent
(ii) The related series of papers in Refs. \cite{Farl1, Farl2, Farl3, Farl4,
Farl5, Farl6, Farl7, Farl8, Farl9}, concerned with evaluating quantum
amplitudes for transitions from initial to final states, in agreement with
a picture where information is not lost, and the end state of black hole
evaporation is a combination of outgoing radiation states.
\vskip 0.3cm
\noindent
(iii) The approach in Refs. \cite{Vilk1, Vilk2, Vilk3}, according to which
black holes create instead a vacuum matter charge to protect themselves
from the quantum evaporation.
\vskip 0.3cm
\noindent
(iv) The work in Ref. \cite{Nico06} where the authors, relying upon the
previous findings in Ref. \cite{Smai04}, consider a noncommutative radiating
Schwarzschild black hole, and find that non-commutativity cures usual problems
encountered in trying to describe the latest stage of black hole evaporation.

We have been therefore led to study how non-commutativity would affect the
analysis of quantum amplitudes in black hole evaporation performed in
Refs. \cite{Farl1, Farl2, Farl3, Farl4, Farl5, Farl6, Farl7, Farl8, Farl9}.
Following Ref. \cite{Nico06}, we assume that non-commutativity of space-time
can be encoded in the commutator of operators corresponding to space-time
coordinates, i.e. (the integer $D$ below is even)
\begin{equation}
\Bigr[x^{\mu},x^{\nu}\Bigr]=i \theta^{\mu \nu}, \;
\mu,\nu=1,2,...,D,
\label{(1)}
\end{equation}
where the antisymmetric matrix $\theta^{\mu \nu}$ is taken to have
block-diagonal form
\begin{equation}
\theta^{\mu \nu}={\rm diag}\Bigr(\theta_{1},...,\theta_{D/2}\Bigr),
\label{(2)}
\end{equation}
with
\begin{equation}
\theta_{i}=\theta
\left(
\begin{array}{cc}
0 & 1 \\
-1 & 0 \\
\end{array}
\right), \;
\forall i =1,2,...,D/2,
\label{(3)}
\end{equation}
the parameter $\theta$ having dimension of length squared and being
constant. As shown in Ref. \cite{Smai04}, the constancy of $\theta$ leads
to a consistent treatment of Lorentz invariance and unitarity.
The authors of Ref. \cite{Nico06} solve the Einstein equations with mass
density of a static, spherically symmetric, smeared particle-like
gravitational source as (hereafter we work in $G=c={\hbar}=1$ units)
\begin{equation}
\rho_{\theta}(r)={M\over (4\pi \theta)^{3\over 2}}e^{-{r^{2}\over 4\theta}},
\label{(4)}
\end{equation}
which therefore plays the role of matter source. Their resulting
spherically symmetric metric is
\begin{eqnarray}
ds^{2}&=&-\left[1-{4M\over r \sqrt{\pi}}\gamma
\left({3\over 2},{r^{2}\over 4\theta}\right)\right]dt^{2}
+\left[1-{4M\over r \sqrt{\pi}}\gamma \left({3\over 2},
{r^{2}\over 4 \theta}\right)\right]^{-1}dr^{2} \nonumber \\
&+& r^{2}(d\Theta^{2}+\sin^{2}\Theta d\phi^{2}),
\label{(5)}
\end{eqnarray}
where we use the lower incomplete gamma function \cite{Nico06}
\begin{equation}
\gamma \left({3\over 2},{r^{2}\over 4\theta}\right) \equiv
\int_{0}^{{r^{2}\over 4\theta}}\sqrt{t}e^{-t}dt.
\label{(6)}
\end{equation}
In this picture, we deal with a mass distribution
\begin{equation}
m(r) \equiv {2M \over \sqrt{\pi}}\gamma \left({3\over 2},
{r^{2}\over 4\theta}\right),
\label{(7)}
\end{equation}
while $M$ is the total mass of the source \cite{Nico06}. This mass
function satisfies the equation
$$
m'(r)=4\pi r^{2} \rho_{\theta}(r),
$$
formally analogous to the general relativity case \cite{Farl9}.

The work in Refs. \cite{Farl1, Farl2, Farl3, Farl4, Farl5, Farl6, Farl7,
Farl8, Farl9} studies instead the quantum-mechanical decay of a
Schwarzschild-like black hole, formed by gravitational collapse, into
almost-flat space-time and weak radiation at a very late time.
The spin-2 gravitational perturbations split into parts with odd and
even parity, and one can isolate suitable variables which can be taken
as boundary data on a final spacelike hypersurface $\Sigma_{F}$. The main
idea is then to consider a complexified classical boundary-value problem
where $T$ is rotated into the complex: $T \rightarrow |T| e^{-i \alpha}$,
for $\alpha \in ]0,\pi/2]$, and evaluate the corresponding classical
Lorentzian action $S_{\rm class}^{(2)}$
to quadratic order in metric perturbations. The
genuinely Lorentzian quantum amplitude is recovered by taking the limit
as $\alpha \rightarrow 0^{+}$ of the semiclassical amplitude
$e^{iS_{\rm class}^{(2)}}$ \cite{Farl1, Farl5, Farl9}.

Section II studies the differential equations obeyed by radial modes
within the framework of the adiabatic approximation, and Sec. III obtains
the resulting orthogonality relation in the presence of a non-vanishing
non-commutativity parameter $\theta$. Section IV derives the effect of
$\theta$ on the expansion of the pure-gravity action functional, which
can be used in the evaluation of quantum amplitudes along the lines of
Refs. \cite{Farl1}-\cite{Farl9}. Absorption and emission spectra of a
non-commutative Schwarzschild-like black hole are studied in Sec. V, while
concluding remarks are presented in Sec. VI, and relevant details are
given in the Appendix.

\section{Equations for radial modes}

The analysis in Ref. \cite{Farl9} holds for any
spherically symmetric Lorentzian background metric
\begin{equation}
ds^{2}=-e^{b(r,t)}dt^{2}+e^{a(r,t)}dr^{2}
+r^{2}(d\Theta^{2}+\sin^{2}\Theta d\phi^{2}),
\label{(8)}
\end{equation}
the even modes $\xi_{2lm}^{(+)}(r,t)$ and odd modes $\xi_{2lm}^{(-)}(r,t)$
being built from a Fourier-type decomposition, i.e. \cite{Farl9}
\begin{equation}
\xi_{2lm}^{(+)}(r,t)=\int_{-\infty}^{\infty}dk \; a_{2klm}^{(+)}
\xi_{2kl}^{(+)}(r){\sin kt \over \sin kT},
\label{(9)}
\end{equation}
and
\begin{equation}
\xi_{2lm}^{(-)}(r,t)=\int_{-\infty}^{\infty}dk \;
a_{2klm}^{(-)}\xi_{2kl}^{(-)}(r){\cos kt \over \sin kT},
\label{(10)}
\end{equation}
where the radial functions $\xi_{2kl}^{(\pm)}$ obey the following
second-order differential equation:
\begin{equation}
e^{-a}{d\over dr}\left(e^{-a}{d\xi_{2kl}^{(\pm)}\over dr}\right)
+\Bigr(k^{2}-V_{l}^{\pm}(r)\Bigr)\xi_{2kl}^{(\pm)}=0,
\label{(11)}
\end{equation}
where, on defining $\lambda \equiv {(l+2)(l-1)\over 2}$, the potential
terms are given by \cite{Farl9}
\begin{equation}
V_{l}^{+}(r)=e^{-a(r,t)}
{2[\lambda^{2}(\lambda+1)r^{3}+3\lambda^{2}mr^{2}+9]m^{2}r+9m^{3}
\over r^{3}(\lambda r+3m)^{2}},
\label{(12)}
\end{equation}
and
\begin{equation}
V_{l}^{-}(r)=e^{-a(r,t)}\left({l(l+1)\over r^{2}}
-{6m\over r^{3}}\right),
\label{(13)}
\end{equation}
respectively. In the expansion of the gravitational action to quadratic
order, it is of crucial importance to evaluate the integral
\begin{equation}
I(k,k',l,R) \equiv \int_{0}^{R}
e^{a(r,t)}\xi_{2kl}^{(+)}(r)\xi_{2k'l}^{(+)}(r)dr,
\label{(14)}
\end{equation}
since \cite{Farl9} (see Appendix)
\begin{equation}
S_{\rm class}^{(2)}[(h_{ij}^{(\pm)})_{lm}]
={\pm 1 \over 32 \pi} \sum_{l=2}^{\infty}\sum_{m=-l}^{l}
{(l-2)! \over (l+2)!}\int_{0}^{R}e^{a}\xi_{2lm}^{(\pm)}
\left({\partial \over \partial t}\xi_{2lm}^{(\pm)*}\right)_{t=T}dr.
\label{(15)}
\end{equation}
For this purpose, we bear in mind the limiting behaviours \cite{Farl9}
\begin{equation}
\xi_{2kl}^{(\pm)} \sim {\rm const} \times (kr)^{l+1}
+{\rm O}((kr)^{l+3}) \; {\rm as} \; r \rightarrow 0,
\label{(16)}
\end{equation}
\begin{equation}
\xi_{2kl}^{(\pm)}(r) \sim z_{2kl}^{(\pm)}e^{ik r_{s}}
+z_{2kl}^{(\pm)*}e^{-ikr_{s}} \; {\rm as} \; r \rightarrow \infty,
\label{(17)}
\end{equation}
where Eq. (16) results from imposing regularity at the origin,
$r_{s}$ is the Regge--Wheeler tortoise coordinate \cite{Farl9, Regg57}
\begin{equation}
r_{s}(r) \equiv r +2M \log(r-2M),
\label{(18)}
\end{equation}
while $z_{2kl}^{(\pm)}$ are complex constants. Indeed, it should be
stressed that non-commutativity can smear plane waves into Gaussian wave
packets. Thus, in a fully self-consistent analysis, the Fourier modes in
Eq. (9), (10), and their asymptotic form in Eq. (16), (17), should be
modified accordingly. However, this task goes beyond the aims of the
present paper, and we hope to be able to perform these calculations in
a future publication.

With this understanding, we can now exploit Eq. (11)
to write the equations (hereafter, we write for simplicity of notation
$\xi_{kl}$ rather than $\xi_{2kl}^{(\pm)}$, and similarly for $V_{l}$
rather than $V_{l}^{\pm}$)
\begin{equation}
e^{a}\xi_{k'l}\left[e^{-a}{d\over dr}\left(e^{-a}{d\over dr}\xi_{kl}\right)
+(k^{2}-V_{l})\xi_{kl}\right]=0,
\label{(19)}
\end{equation}
\begin{equation}
e^{a}\xi_{kl}\left[e^{-a}{d\over dr}\left(e^{-a}{d\over dr}\xi_{k'l}\right)
+({k'}^{2}-V_{l})\xi_{k'l}\right]=0.
\label{(20)}
\end{equation}
According to a standard procedure, if we subtract Eq. (20) from Eq. (19),
and integrate the resulting equation from $r=0$ to $r=R$, we obtain
\begin{equation}
(k^{2}-{k'}^{2})\int_{0}^{R}e^{a}\xi_{kl}\xi_{k'l}dr
= \int_{0}^{R}\left[\xi_{kl}{d\over dr}\left(e^{-a}{d\over dr}
\xi_{k'l}\right)-\xi_{k'l}{d\over dr}\left(e^{-a}{d\over dr}\xi_{kl}
\right)\right]dr.
\label{(21)}
\end{equation}
The desired integral (14) is therefore obtained from Eq. (21),
whose right-hand side is then
completely determined from the limiting behaviours in
Eqs. (16) and (17), i.e.
\begin{equation}
\int_{0}^{R}e^{a}\xi_{kl}\xi_{k'l}dr=
\left \{ {1\over (k^{2}-{k'}^{2})}\left[\xi_{kl}e^{-a}
\left({d\over dr}\xi_{k'l}\right)-\xi_{k'l}e^{-a}
\left({d\over dr}\xi_{kl}\right)\right]_{r=0}^{r=R}
\right \},
\label{(22)}
\end{equation}
where, on going from Eq. (21) to Eq. (22), we have exploited the
vanishing coefficient that weights the integral
$$
\int_{0}^{R}e^{-a}\left({d\over dr}\xi_{kl}\right)
\left({d\over dr}\xi_{k'l}\right)dr,
$$
resulting from two contributions of equal magnitude and opposite sign.
By virtue of Eq. (16), $r=0$ gives vanishing contribution to the right-hand
side of Eq. (22), while the contribution of first derivatives of radial
functions involves also
\begin{equation}
\left . {dr_{s}\over dr} \right |_{r=R}
={R\over (R-2M)}.
\label{(23)}
\end{equation}

\section{Generalized orthogonality relation}

Note now that our metric (5) is a particular case of the spherically
symmetric metric (8), since our $a$ and $b$ functions are independent
of time. More precisely, unlike the full Vaidya space-time, where in the
region containing outgoing radiation the mass function varies extremely
slowly with respect both to $t$ and to $r$
\cite{Farl8}, we consider a ``hybrid''
scheme where the mass function depends on $r$ only for any fixed value of
the non-commutativity parameter $\theta$. We can therefore write
\begin{equation}
e^{-a}=1-{4M\over r \sqrt{\pi}}
\gamma \left({3\over 2},{r^{2}\over 4\theta}\right)
\label{(24)}
\end{equation}
in our non-commutative spherically symmetric model,
where the function in curly brackets in Eq. (22) reads as
$$
\left(1-{4M \over R \sqrt{\pi}}\gamma \left({3\over 2},
{R^{2}\over 4\theta}\right)\right){R \over (R-2M)}
{1\over (k^{2}-{k'}^{2})}
\times i \Bigr[(k'-k)z_{kl}z_{k'l}e^{i(k+k')r_{s}(R)}
$$
$$
+(k-k')z_{kl}^{*}z_{k'l}^{*}e^{-i(k+k')r_{s}(R)}
-(k+k')z_{kl}z_{k'l}^{*}e^{i(k-k')r_{s}(R)}
+(k+k')z_{kl}^{*}z_{k'l}e^{i(k'-k)r_{s}(R)}\Bigr].
$$
At this stage, we exploit one of the familiar limits that can be used
to express the Dirac $\delta$, i.e. \cite{Farl9}
\begin{equation}
\lim_{r_{s}\to \infty}{e^{i(k \pm k')r_{s}}\over (k \pm k')}
=i \pi \delta(k \pm k'),
\label{(25)}
\end{equation}
to find
\begin{equation}
\int_{0}^{R}e^{a}\xi_{kl}\xi_{k'l}dr=2\pi |z_{kl}|^{2}
F(R,\theta) \Bigr(\delta(k+k')+\delta(k-k')\Bigr),
\label{(26)}
\end{equation}
having defined
\begin{equation}
F(R,\theta) \equiv {R \over (R-2M)} \left[1-
{4M \over R \sqrt{\pi}}\gamma \left({3\over 2},{R^{2}\over 4\theta}\right)
\right].
\label{(27)}
\end{equation}

\section{Effect of $\theta$ and expansion of the action functional}

Since $\theta$ has dimension length squared as we said after Eq. (3),
we can define the noncommutativity-induced length scale
\begin{equation}
L \equiv 2 \sqrt{\theta}.
\label{(28)}
\end{equation}
Moreover, we know that
our results only hold in the adiabatic approximation, i.e. when both
$m'$ and ${\dot m}$ are very small. The latter condition is obviously
satisfied because our mass function in Eq. (7) is independent of time.
The former amounts to requiring that (hereafter we set
$w \equiv R/L$, while $R_{s} \equiv 2M$)
\begin{equation}
m'(R)={2\over \sqrt{\pi}}{R_{s}\over L}e^{-w^{2}}w^{2} <<1.
\label{(29)}
\end{equation}
The condition (29) is satisfied provided that either
\vskip 0.3cm
\noindent
(i) $w \rightarrow \infty$ or $w \rightarrow 0$, i.e. $R >>L$ or
$R << L$;
\vskip 0.3cm
\noindent
(ii) or at $R=L$ such that
\begin{equation}
m'(R=L)=m'(w=1)={2\over \sqrt{\pi}}{R_{s}\over L}e^{-1} <<1,
\label{(30)}
\end{equation}
and hence for ${R_{s}\over L} << {e \sqrt{\pi} \over 2}$.

Furthermore, at finite values of the non-commutativity parameter $\theta$,
our $w \equiv {R\over L}$ is always much larger than $1$ in Eq. (27)
if $R$ is very large, and hence
we can exploit the asymptotic expansion of the lower incomplete
$\gamma$-function in this limit \cite{Abra64, Spal06}, i.e.
\begin{equation}
\gamma \left({3\over 2},w^{2}\right)=
\Gamma \left ({3\over 2} \right)
-\Gamma \left({3\over 2},w^{2}\right)
\sim {1\over 2}\sqrt{\pi}\left[1-e^{-w^{2}}
\sum_{p=0}^{\infty}{w^{1-2p}
\over \Gamma \left({3\over 2}-p \right)}\right].
\label{(31)}
\end{equation}
By virtue of Eqs. (27) and (31), we find
\begin{equation}
F(R,\theta) \equiv F(R,L) \sim 1+{R_{s} \over (R-R_{s})}
e^{-w^{2}}
\sum_{p=0}^{\infty}{w^{1-2p}\over
\Gamma \left({3\over 2}-p \right)}.
\label{(32)}
\end{equation}
Equation (32) describes the asymptotic expansion
of the correction factor $F$ when $R >> L$.

In the opposite regime, i.e. for $\theta$ so large that
$(R/L)<<1$ despite that $R$ tends to $\infty$, one has
\cite{Spal06}
\begin{equation}
F(R,L) \sim {R \over (R-R_{s})}\left[
1-{4 \over 3 \sqrt{\pi}}{R_{s}\over R}w^{3}
\left(1-{7\over 5}w^{2}\right)\right].
\label{(33)}
\end{equation}

Last, but not least, if $R$ and $L$ are comparable, the lower-incomplete
$\gamma$-function in Eq. (27) cannot be expanded, and we find,
bearing in mind that $R_{s}/L <<1$ from Eq. (30), the limiting form
\begin{equation}
F(R,L) \sim 1+ {R_{s}\over L} \left(1-{2\over \sqrt{\pi}}
\gamma \left({3\over 2},1 \right)\right)
+{\rm O}((R_{s}/L)^{2}).
\label{(34)}
\end{equation}

We therefore conclude that a $\theta$-dependent correction to the
general relativity analysis in Ref. \cite{Farl9} does indeed arise from
non-commutative geometry. In particular, the expansion of the
action to quadratic order in perturbative modes takes the form
(cf. Ref. \cite{Farl9})
\begin{equation}
S_{\rm class}^{(2)}={F(R,L)\over 16} \sum_{l=2}^{\infty}
\sum_{m=-l}^{l}\sum_{P=\pm 1}{(l-2)!\over (l+2)!}\int_{0}^{\infty}dk \;
k |z_{2klP}|^{2} \left | a_{2klmP}+P a_{2,-klmP} \right |^{2}
\cot kT,
\label{(35)}
\end{equation}
where the function $F(R,L)$ (see Eq. (27)) takes the
limiting forms (32) and (33), respectively, depending on whether
$w>>1$ or $w<<1$, while
$P = \pm 1$ for even (respectively odd) metric perturbations. Our
``correction'' $F(R,L)$ to the general relativity analysis is
non-vanishing provided that one works at very large but finite values
of $R$. In the limit as $R \rightarrow \infty$, one has instead
\begin{equation}
\lim_{R \to \infty} F(R,L)=1,
\label{(36)}
\end{equation}
which means that, at infinite distance from the Lorentzian singularity
of Schwarzschild geometry, one cannot detect the effect of a
non-commutativity parameter.

\section{Absorption and emission spectra of a non-commutative
Schwarzschild black hole}

If general relativity is taken as the fundamental description of
gravitational phenomena, particle emission by a black hole follows
a Planck-type spectrum \cite{Hawk75} while wave absorption by a black
hole shows an oscillatory behaviour as a function of frequency
\cite{Sanc78}. For a Schwarzschild black hole, the two spectra are
related by
\begin{equation}
H(\omega)={\sigma (\omega)\over e^{\omega/T_H}-1}, 
\label{(37)}
\end{equation}
where $\omega$ is the frequency of the wave and $T_H$ is the
temperature for a body emitting thermal radiation, i.e. for the
black hole. With this notation, $H(\omega)$ and $\sigma(\omega)$
are the emission and absorption rates, respectively. In Ref.
\cite{Sanc78} the author has found the following expression for
the total absorption cross-section $\sigma(\omega)$ in the Hawking
formula (37):
\begin{equation}
\sigma(\omega)=27 \pi M^{2}-2 \sqrt{2}M
{\sin(2 \sqrt{27} \pi \omega M)\over \omega},
\label{(38)}
\end{equation}
where $\sigma(\omega)$ oscillates about the value $27 \pi M^{2}$ with
decreasing amplitude ${2 \sqrt{2} M \over \omega}$ and an approximately
constant oscillation period.

We can `import' the above result in our case by virtue of the
adiabatic approximation exploited in Refs.
\cite{Farl1}-\cite{Farl9}, and we hence replace $T_H$ in
Eq. (\ref{(37)}) by the noncommutative black hole temperature
$T_{HNC}$ defined as
\begin{equation}
T_{HNC}\equiv
- \left(\frac{1}{4\pi}\frac{dg_{00}}{dr}\right)_{r=r_+}=
\frac{1}{4\pi
r_+}\left[1-\frac{r_+^3}{4\theta^{3/2}}
\frac{e^{-r_+^2/4\theta}}{\gamma(3/2;r_+^2/4\theta)}\right],
\label{(39)}
\end{equation}
where $r_+$ solves the horizon equation
\begin{equation}
r_+=2m(r_+,\theta) \equiv {4M \over \sqrt{\pi}}\gamma
\left({3\over 2}, {r_+^{2}\over 4\theta}\right). 
\label{(40)}
\end{equation}
Equation (38) yields therefore
\begin{equation}
\sigma(\omega)=\frac{27 \pi}{4} r_+^{2}- \sqrt{2}r_+ {\sin(
\sqrt{27} \pi \omega r_+)\over \omega}, 
\label{(41)}
\end{equation}
where $\sigma(\omega)$ oscillates about ${27\over 4} \pi r_+^{2}$ with
decreasing amplitude ${ \sqrt{2}r_+\over \omega}$ and a smaller
oscillation period depending on $\theta$ (note that
the $\theta$-dependence in the horizon radius
dominates at small $\theta$). Moreover, the emission spectrum of
the black hole takes the form
\begin{equation}
H(\omega)={\sigma(\omega) \over e^{ \omega/T_{HNC}}-1 }.
\label{(42)}
\end{equation}
In the `large radius' regime $r_+^{2}/4\theta >> 1$, i.e. $R>>L$,
Eq. (\ref{(40)}) can be solved by iteration \cite{Nico06}. To
first order in $M/\sqrt{\theta}$, one finds
\begin{equation}
r_+=2M\left(1-{M \over \sqrt{\pi\theta}}e^{-M^2/\theta}\right) .
\label{(43)}
\end{equation}
The resulting plot of emission spectrum of the black hole
(\ref{(42)}) in the `large radius' regime, i.e. if $R
>>L$ ($M >>L$) is shown in Fig. 1. 
In Fig. 1 the non-commutativity parameter in
Eq. (28) is unable to modify the general relativity shape, although
the maximum value attained is smaller than in general relativity.

\begin{figure}[!h]
\centerline{\hbox{\psfig{figure=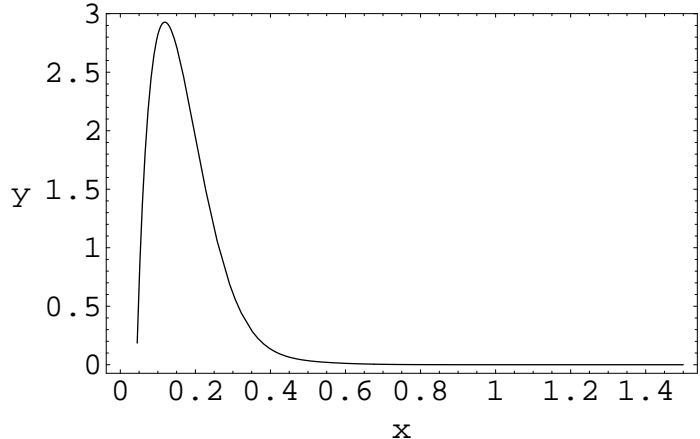,width=0.6\textwidth}}}
\caption{Total emission spectrum for a black hole in units $y
\equiv H(x)/4M^{2}$, where $x \equiv 2M \omega$ and $R=10L$.
\label{fig1}}
\end{figure}

\section{Concluding remarks}

Our paper has investigated the effect of non-commutative geometry on the
recent theoretical analysis of quantum amplitudes in black hole
evaporation, following the work in Refs. \cite{Hawk05},
\cite{Farl1}-\cite{Farl9} (for other developments, see for example the
recent work in Refs. \cite{Kar05, Kar06a, Kar06b, Albe06}).
For this purpose, we have considered an
approximate scheme where the background space-time is static and
spherically symmetric, with mass function depending on the radial
coordinate only for any fixed value of the non-commutativity parameter
$\theta$.

Within this framework, we find that the general relativity analysis
of spin-2 amplitudes is modified by a multiplicative factor $F$ defined
in Eq. (27). Its limiting forms for $R >>L$ or $R <<L$ or
$R \cong L$ are given by Eqs. (32), (33) and (34), respectively.
Within this framework, unitarity is preserved, and the end state of
black hole evaporation is a combination of outgoing radiation states
(see section 1).

When the adiabatic approximation here assumed holds, we have also
considered approximate formulae for the absorption and emission
spectra of a ``non-commutative Schwarzschild'' black hole. The
resulting plot shows that for the total emission spectrum the
General Relativity shape is essentially recovered at finite values
of $\theta$ such that $R >> L$. Within such a scheme, Hawking
emission is only important in a certain frequency range
\cite{Sanc78}, but nothing can be said about the end-state of
black hole evaporation.

An outstanding open problem is
whether one can derive a time-dependent spherically
symmetric background metric which incorporates the effects of
non-commutative geometry. This would make it possible to improve the
present comparison with the results in Refs. \cite{Farl1}-\cite{Farl9},
where the Vaidya space-time was taken as the background geometry.
A closer inspection of the effect of a variable surface gravity is also
in order, jointly with an assessment of the whole subject of black hole
thermodynamics, when ``corrected'' by non-commutative geometry.

\acknowledgments
The work of G. Esposito and G. Miele has been partially supported by
PRIN {\it SINTESI} and PRIN {\it FISICA ASTROPARTICELLARE},
respectively.

\appendix
\section{From Regge--Wheeler to asymptotically flat gauge}

Although we refer the reader to Ref. \cite{Farl9} for the large amount
of detailed calculations, we should stress a few important points about
the derivation of Eq. (15). The metric perturbations of linearized theory
are subject to infinitesimal diffeomorphisms, which are the
`gauge transformations' of general relativity \cite{Dewi03} (since the
infinite-dimensional invariance group of the Einstein theory is indeed
the diffeomorphism group).

The Regge--Wheeler (hereafter RW) gauge is not a supplementary condition,
but rather an infinitesimal diffeomorphism according to which the odd-parity
metric perturbations are modified as follows:
\begin{equation}
h_{0lm}^{(-)RW}=h_{0lm}^{(-)'}=h_{0lm}^{(-)}-\partial_{t}
\Lambda_{lm},
\label{(A1)}
\end{equation}
\begin{equation}
h_{1lm}^{(-)RW}=h_{1lm}^{(-)'}=h_{1lm}^{(-)}
-\partial_{r}\Lambda_{lm}+{2\Lambda_{lm}\over r},
\label{(A2)}
\end{equation}
\begin{equation}
h_{2lm}^{(-)RW}=0=h_{2lm}^{(-)'}=h_{2lm}^{(-)}+2\Lambda_{lm},
\label{(A3)}
\end{equation}
where $h_{0lm}^{(-)}$ occurs in the expansion of the shift vector, while
$h_{1lm}^{(-)}$ and $h_{2lm}^{(-)}$ occur in the expansion of the
odd-parity three-metric perturbations \cite{Farl9}.

In our spherically symmetric model of non-commutative gravity, since we
rely upon the work in Ref. \cite{Nico06}, where the left-hand side of the
Einstein equations retains the same functional form as in general
relativity, it remains possible to consider infinitesimal diffeomorphisms
formally analogous to Eqs. (A1)--(A3). For a more general framework, that
we do not strictly need here, one should instead build a deformation of
the algebra of diffeomorphisms, e.g. along the lines of the work in
Refs. \cite{Asch05, Asch06}.

In the RW gauge, one of the coupled partial differential equations relating
$h_{1lm}$ and $h_{0lm}$ is \cite{Farl9}
\begin{equation}
\partial_{t}^{2} h_{1lm}^{RW}=\partial_{r}\partial_{t}
h_{0lm}^{RW}-{2\over r}\partial_{t}h_{0lm}^{RW}
+\left[-{2\lambda e^{b}\over r^{2}}-{2e^{b}\over r}
\left(m''+{2m'e^{a}\over r^{2}}(m'+rm)\right) \right]h_{1lm}^{RW}.
\label{(A4)}
\end{equation}
To recover the expected fall-off behaviour of metric perturbations, one
later performs the asymptotically flat (hereafter AF) gauge transformation,
according to which \cite{Farl9}
\begin{equation}
h_{0lm}^{(-)AF}=0=h_{0lm}^{(-)RW}-\partial_{t}\Lambda_{lm},
\label{(A5)}
\end{equation}
\begin{equation}
h_{1lm}^{(-)AF}=h_{1lm}^{(-)RW}-\partial_{r}\Lambda_{lm}
+{2\Lambda_{lm}\over r},
\label{(A6)}
\end{equation}
\begin{equation}
h_{2lm}^{(-)AF}=h_{2lm}^{(-)RW}+2\Lambda_{lm}=2\Lambda_{lm}.
\label{(A7)}
\end{equation}
By virtue of Eqs. (A6) and (A4), and neglecting all derivatives of the
mass function (this remains legitimate for our mass function in
Eq. (7), as is clear from Fig. 2 below),
one finds
\begin{equation}
\partial_{t}^{2}h_{1lm}^{(-)AF}=-{2\lambda e^{b}\over r^{2}}
h_{1lm}^{(-)RW},
\label{(A8)}
\end{equation}
since exact cancellations occur of the coefficients of
$\partial_{r}\partial_{t}^{2}\Lambda_{lm}$ and
$\partial_{t}^{2}\Lambda_{lm}$. Moreover, since $h_{2lm}^{(-)}$
vanishes in the RW gauge (see Eq. (A3)), one finds, for the Zerilli
function \cite{Farl9, Zeri70},
\begin{equation}
Q_{lm}^{(-)RW} \equiv {e^{-a}\over r}\left(h_{1lm}^{(-)RW}
+{r^{2}\over 2}\partial_{r}{h_{2lm}^{(-)RW}\over r}\right)
={e^{-a}\over r}h_{1lm}^{(-)RW}.
\label{(A9)}
\end{equation}
With our metric (5), one has $e^{b}=e^{-a}$, and hence
\begin{equation}
\partial_{t}^{2}h_{1lm}^{(-)AF}=-{2\lambda \over r}Q_{lm}^{(-)RW},
\label{(A10)}
\end{equation}
in complete analogy with the general relativity analysis in Ref.
\cite{Farl9}. Equation (A10) is in turn used to prove the desired
fall-off property
\begin{equation}
h_{1lm}^{(-)AF}(r,t)={2\lambda \over r}\int_{-\infty}^{\infty}
dk \; {a_{klm}^{(-)}\over k^{2}}Q_{kl}^{(-)RW}
{\sin kt \over \sin kT},
\label{(A11)}
\end{equation}
after writing \cite{Farl9}
\begin{equation}
Q_{lm}^{(-)RW}=\int_{-\infty}^{\infty}dk \; a_{klm}^{(-)}
Q_{kl}^{(-)RW}{\sin kt \over \sin kT}.
\label{(A12)}
\end{equation}

\begin{figure}[!h]
\centerline{\hbox{\psfig{figure=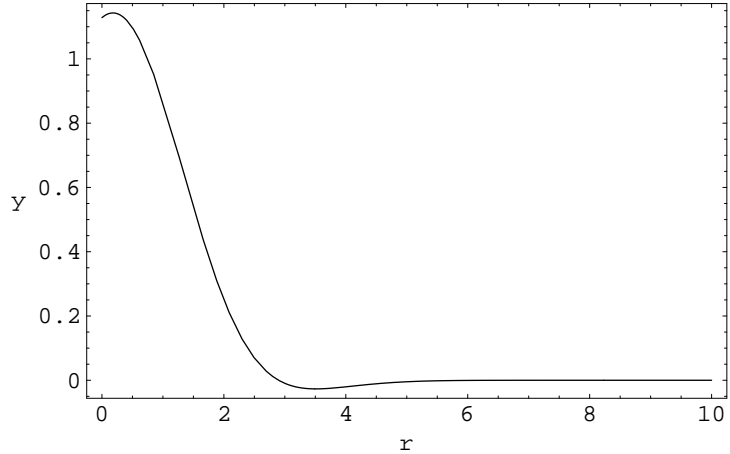,width=0.6\textwidth}}}
\caption{Plot of the function $y \equiv m''+2m'e^{a}(m'+rm)r^{-2}$,
with $m$ defined as in Eq. (7) and $\theta=1$ (hence much smaller
than $R^{2}$).
\label{fig2}}
\end{figure}

\end{document}